\documentclass[conference]{IEEEtran}
\IEEEoverridecommandlockouts
\usepackage{cite}
\usepackage{amsmath,amssymb,amsfonts}
\usepackage{pifont}
\usepackage{algorithmic}
\usepackage{graphicx}
\usepackage{url}
\usepackage{booktabs}
\usepackage{float}
\usepackage{multirow}
\usepackage{textcomp}
\usepackage{makecell}
\usepackage{xcolor}
\usepackage{array}
\usepackage{todonotes}
\usepackage{threeparttable}
\def\BibTeX{{\rm B\kern-.05em{\sc i\kern-.025em b}\kern-.08em
    T\kern-.1667em\lower.7ex\hbox{E}\kern-.125emX}}
    
\newcommand{\xmark}{\ding{55}}%

\begin{document}

\title{Demystifying NCCL: An In-depth Analysis of GPU Communication Protocols and Algorithms}

\author{
\IEEEauthorblockN{
Zhiyi Hu$^1$\IEEEauthorrefmark{1},
Siyuan Shen$^1$\IEEEauthorrefmark{1},
Tommaso Bonato$^1$,
Sylvain Jeaugey$^2$,
Cedell Alexander$^3$,\\
Eric Spada$^3$,
James Dinan$^2$,
Jeff Hammond$^2$,
Torsten Hoefler$^1$
}

\IEEEauthorblockA{$^1$ETH Zürich, Switzerland}
\IEEEauthorblockA{zhiyihu@student.ethz.ch, \{siyuan.shen, tommaso.bonato, torsten.hoefler\}@inf.ethz.ch}

\IEEEauthorblockA{$^2$NVIDIA Corporation, \{sjeaugey, jdinan, jehammond\}@nvidia.com}

\IEEEauthorblockA{$^3$Broadcom Inc., \{cedell.alexander, eric.spada\}@broadcom.com}
\IEEEauthorblockA{* These authors contributed equally to this work}
}

\maketitle

\begin{abstract}
The NVIDIA Collective Communication Library (NCCL) is a critical software layer enabling high-performance collectives on large-scale GPU clusters. Despite being open source with a documented API, its internal design remains largely opaque. The orchestration of communication channels, selection of protocols, and handling of memory movement across devices and nodes are not well understood, making it difficult to analyze performance or identify bottlenecks. This paper presents a comprehensive analysis of NCCL, focusing on its communication protocol variants (Simple, LL, and LL128), mechanisms governing intra-node and inter-node data movement, and ring- and tree-based collective communication algorithms. The insights obtained from this study serve as the foundation for ATLAHS, an application-trace-driven network simulation toolchain capable of accurately reproducing NCCL communication patterns in large-scale AI training workloads. By demystifying NCCL’s internal architecture, this work provides guidance for system researchers and performance engineers working to optimize or simulate collective communication at scale.
\end{abstract}

\begin{IEEEkeywords}
NVIDIA NCCL, Collective communication, Communication libraries, Multi-GPU cluster training
\end{IEEEkeywords}

\section{Introduction}
Efficient GPU-to-GPU communication is essential for achieving high performance in distributed artificial intelligence (AI) and high-performance computing (HPC) workloads. The NVIDIA Collective Communication Library (NCCL) is a prominent library widely adopted for scalable, optimized GPU communication~\cite{ncclgithub, nccl_documentation}. Unlike general-purpose message-passing frameworks, such as MPI~\cite{mpi}, NCCL specifically targets GPU-to-GPU interactions, utilizing interconnect technologies such as NVLink, PCIe, and InfiniBand (IB) to achieve high bandwidth and low latency.

Although NCCL is critical to large-scale GPU systems and open source, its internal mechanisms remain insufficiently documented. This is reflected by the frequent technical questions posted on NCCL’s GitHub page~\cite{ncclgithub}, where users seek details about the library’s inner workings. While the official API documentation is thorough, key aspects such as topology construction, algorithm selection, pipelining, and buffer management across nodes and devices are not clearly described. This lack of transparency makes it difficult for system researchers, network architects, and performance engineers to optimize, or predict NCCL’s performance on new hardware and at scale~\cite{exploring_gpu_to_gpu_desensi, understanding_gh200_fusco}.

In this paper, we present a thorough and systematic exploration of NCCL’s internal architecture. Our analysis specifically targets four primary aspects of NCCL’s implementation: (1) a general overview, including API structure and communication channel management; (2) a detailed examination of communication protocols (Simple, LL, LL128); (3) an analysis of its data-transfer models; and (4) comprehensive analysis of its collective communication algorithms.

The insights gained from this study provide important context for performance modeling and architectural optimization. These insights have been adopted in simulation frameworks such as ATLAHS~\cite{atlash_shen}, an application-trace-driven network simulator developed to accurately replicate the communication patterns of NCCL-based machine learning workloads. By clarifying NCCL’s internal design principles, this analysis supports system researchers, interconnect designers, and network architects in making more informed optimization decisions for GPU-centric high-performance computing environments.

The analysis in this paper is based on NCCL version 2.19.1. While specific implementation details may evolve in future releases, the core architectural mechanisms and communication strategies discussed here are expected to remain consistent, ensuring that the insights presented remain broadly applicable.

\section{NCCL Overview}

\subsection{NCCL API}

NCCL is specifically designed to provide highly optimized collective communication operations for GPU clusters, emphasizing low latency and high bandwidth. At its core, NCCL manages GPU-to-GPU communication via a clear and efficient API that abstracts complex technical details. NCCL primarily provides four categories of functions to users:

\subsubsection{Communicator Management}

Similar to MPI, all communication operations in NCCL are performed within the context of \textbf{communicators}. Each GPU participating in communication maintains a communicator object, which is used to invoke NCCL operations. Users must first initialize a communicator and define the set of GPUs involved.

When all devices are managed within a single process or thread, \texttt{ncclCommInitAll} can be used to create the communicator collectively. For multi-process or multi-threaded environments, each process calls \texttt{ncclCommInitRank} with a shared unique identifier to correctly establish the communicator across processes.

After communication tasks have finished, communicators should be properly released to free resources. NCCL provides two functions for this purpose:

\begin{itemize}
\item \texttt{ncclCommDestroy}: Safely destroys a communicator, ensuring all pending communication operations are completed before cleanup.
\item \texttt{ncclCommAbort}: Immediately terminates the communicator and cancels ongoing operations. This is intended for error recovery or handling unexpected failures to avoid deadlocks.
\end{itemize}

\subsubsection{Collective Communication}

NCCL provides 5 collective operations: \texttt{ncclAllReduce}, \texttt{ncclBroadcast}, \texttt{ncclReduce}, \texttt{ncclAllGather}, and \texttt{ncclReduceScatter}. Historically, NCCL included an in-place variant of \texttt{ncclBroadcast}, called \texttt{ncclBcast}, to mimic the behavior of \texttt{MPI\_Bcast}, which always operates in-place. However, to support more general use cases and achieve a more regular API, NCCL later introduced \texttt{ncclBroadcast} with separate send and receive buffers. \texttt{ncclBcast} is now largely deprecated and maintained primarily for compatibility with MPI-style interfaces.

\subsubsection{Point-to-Point Communication}

NCCL supports point-to-point operations through \texttt{ncclSend} and \texttt{ncclRecv}.

\subsubsection{Group Calls}
To aggregate operations and reduce overhead, NCCL offers \texttt{ncclGroupStart} and \texttt{ncclGroupEnd}. These functions bracket a sequence of NCCL calls and delay their execution until the group ends. Grouped operations may include multiple Send/Recv calls (to emulate SendRecv, All-to-One, One-to-All, or All-to-All patterns) or a set of collective operations. This aggregation can significantly reduces launch overhead and latency by ensuring that all grouped operations are executed together as part of a single NCCL launch.

\subsection{Launching Strategies}
\label{sec:task-launch-strategies}

NCCL supports three common execution models for launching operations on multiple GPUs, and each of these approaches presents distinct trade-offs. 

\begin{itemize}
    \item \textbf{One CPU process per GPU:} This model provides greater control over process placement. By binding each GPU to a separate process, the associated CPU code can be scheduled on the local non-uniform memory access (NUMA) domain, improving data locality and reducing memory access latency.
    
    \item \textbf{One CPU thread per GPU:} When a single CPU process manages multiple GPUs through multiple threads, it enables efficient intra-process memory sharing. This setup allows for direct access to memory across ranks, including GPU buffers, reducing memory-copy overhead during communication.
    
    \item \textbf{One CPU thread for multiple GPUs:} While the single-threaded model suffers from sequential kernel launches and reduced concurrency, it offers simplicity, minimal CPU overhead, and deterministic execution, making it suitable for small-scale deployments or prototype environments where ease of implementation is prioritized over the highest performance.
\end{itemize}

\subsection{Communication Channels}

NCCL orchestrates communication through three hardware components: the GPU, the CPU, and the network interface. GPUs execute reductions and move data between buffers, CPUs launch kernels and manage host-side coordination, and NICs transfer packets across nodes. When only a single streaming multiprocessor (SM) handles the GPU work, large messages can overload that SM, underuse other SMs, and fail to saturate links such as NVLink or InfiniBand~\cite{nccl_issue_578, nccl_issue_1302}.

To avoid this bottleneck, NCCL subdivides every collective into \textbf{communication channels}. Each channel is launched as a separate CUDA block that runs on its own SM, and the library partitions the input buffer so that channels operate on disjoint chunks in parallel. This fine-grained parallelism raises aggregate throughput, especially for large payloads that would otherwise serialize on one SM. Spreading work across channels also helps balance traffic across multiple NICs on NVLink platforms, as each channel can independently exit the node through a different NIC. This improves link utilization, reduces idle time, and balances load across interconnects such as NVLink, PCIe, and InfiniBand.

However, aggressive use of multiple channels can negatively impact network efficiency. When the per-channel chunk size becomes smaller than the 512 KiB FIFO buffer size employed by NIC transports, the proxy thread sends partially filled buffers. This under-utilization can degrade PCIe and network throughput, particularly when multiple queue pairs (QPs) are active to enable Equal-Cost Multi-Path Routing (ECMP) load balancing. NCCL addresses this issue by heuristically reducing \texttt{nChannels} for smaller messages (refer to the function \texttt{calcP2pChunkSize} in \texttt{enqueue.cc}). Nevertheless, selecting an optimal channel count remains a trade-off between GPU-side parallelism and network utilization efficiency.

Channel management in NCCL is coordinated at the communicator level, where each GPU receives a unique rank between $0$ and $n-1$, where $n$ is the total number of GPUs participating in the communicator. During communicator initialization, NCCL establishes an initial set of channel structures, with their total count primarily guided by system topology and architectural defaults. When a collective operation is invoked, NCCL dynamically selects the algorithm and protocol for that particular task. Based on this runtime choice, NCCL's internal tuning model then determines how many of these pre-established channels to utilize for that operation, considering the selected strategy, current message size, available bandwidth, and configured threads per channel. Although earlier versions allowed users to influence channel behavior by setting environment variables like \texttt{NCCL\_NTHREADS}, such manual tuning is now discouraged. In recent versions, these settings are typically ignored and may even lead to incorrect behavior.

The logical communication topology assigned to each channel directly shapes how data flows among GPUs during each operation. In a \textbf{ring topology}, each GPU identifies its immediate predecessor and successor to form a unidirectional communication ring. In a \textbf{tree topology}, each GPU tracks its parent and child ranks, establishing a logical communication tree. To increase bandwidth utilization, NCCL employs a \textbf{double binary tree} structure~\cite{dt, colls}: no node is a non-leaf in both trees, and at most one node appears as a leaf in both. The second tree is constructed by mirroring the first when the number of nodes is even, or by a one-position shift when it is odd. These topologies are established during communicator initialization and reused across all collective operations.

For grouped point-to-point operations using \texttt{ncclGroupStart} and \texttt{ncclGroupEnd}, NCCL assigns each transfer to a separate channel when possible, enabling multiple independent sends and receives to run in parallel. This provides task-level parallelism across transfers. %

\section{Communication Protocols}

NCCL employs multiple communication protocols to optimize data transfer efficiency during collective operations. The three protocols, \textbf{Simple}, \textbf{LL} (Low Latency), and \textbf{LL128}, are designed to achieve different trade-offs between bandwidth and latency. This section provides an overview of the mechanisms behind each protocol. Table~\ref{tab:protocol_comparison} summarizes the key characteristics of the three protocols.

\begin{table}[!htp]
    \centering
    \caption{Comparison of NCCL Communication Protocols}
    \label{tab:protocol_comparison}
    \renewcommand{\arraystretch}{1.5}
    \begin{tabular}{@{}cccc@{}}
        \toprule
        & \textbf{Simple} & \textbf{LL} & \textbf{LL128} \\
        \midrule
        \textbf{Design Goal} & High bandwidth & Low latency & \makecell{Low latency and\\high bandwidth} \\
        \midrule
        \makecell{\textbf{Synchronization} \\ \textbf{Mechanism}} & \makecell{Memory fences\\(high overhead)} & \makecell{Flag-based\\synchronization} & \makecell{Flag-based\\synchronization} \\
        \midrule
        \textbf{Payload} & Data chunks & \makecell{4B data +\\4B flag} & \makecell{120B data +\\8B flag} \\
        \midrule
        \makecell{\textbf{Bandwidth} \\ \textbf{Utilization}} & Near peak & \makecell{$25\sim50\%$\\of peak~\cite{nccl_jeaugey}} & \makecell{$\sim95\%$\\of peak~\cite{nccl_jeaugey}} \\
        \midrule
        \textbf{Latency Per-hop} & $\sim 6\mu s$ & $\sim 1 \mu s$ & $\sim 2 \mu s$ \\
        \bottomrule
    \end{tabular}
    \vspace{-1em}
\end{table}

\subsection{Simple Protocol}

The \textit{Simple protocol} is designed to maximize bandwidth utilization and is used for large message transfers. It operates by dividing the data into relatively large chunks and dispatching them across communication channels. This chunking strategy ensures that the high throughput of the network interface and GPU memory system is fully leveraged.

To preserve memory consistency, the protocol uses memory fences to enforce correct ordering and visibility of data. A receiver must wait until a full chunk has been transferred before accessing it. While effective at ensuring correctness, the use of memory fences introduces significant overhead. This overhead becomes a limiting factor for small messages, where the cost of synchronization dominates overall transfer time. As a result, while the Simple protocol achieves near-peak bandwidth for large messages, it suffers from high latency when handling small payloads.

\subsection{LL (Low Latency) Protocol}

To address the latency issues associated with the Simple protocol, NCCL includes the \textit{LL protocol}, which is optimized for small message sizes where bandwidth is typically underutilized. Instead of relying on memory fences, the LL protocol uses lightweight flag-based synchronization. A small flag is transmitted alongside the data to signal its validity, enabling the receiver to proceed as soon as the data becomes available without requiring costly memory barriers.

Each transmission in the LL protocol consists of 4 bytes of data followed by a 4-byte flag, sent together using 8-byte atomic operations. This approach significantly reduces synchronization overhead and improves responsiveness for latency-sensitive workloads. LL forces the intermediate buffer to reside in host memory so that the CPU can poll the flag and detect when the data is ready to be sent through the NIC.
This is necessary because polling GPU memory over PCIe is much slower than DRAM access and requires explicit synchronization to ensure data visibility on the host. While this design enables low latency, it prevents the use of GPU Direct Remote Direct Memory Access (RDMA), severely limiting bandwidth. As a result, LL typically achieves only 25–50 percent of peak bandwidth, depending on the interconnect. Consequently, it is preferred only for small transfers where latency is critical and bandwidth utilization is secondary.

\subsection{LL128 Protocol}

The \textit{LL128 protocol} improves upon LL by maintaining its low-latency properties while significantly increasing bandwidth efficiency, particularly over high-performance interconnects like NVLink. Like LL, it uses flag-based synchronization to eliminate memory fences, but it transmits data in 128-byte units rather than 8-byte units. Out of the 128 bytes, 120 bytes are dedicated to data, and 8 bytes are reserved for the flag, allowing the protocol to utilize approximately 95 percent of the peak bandwidth.

On the network path, LL128 resembles the Simple protocol in that the sending GPU aggregates a relatively large chunk of data before notifying the CPU that it is ready to send. Although this chunk-based aggregation limits pipelining across nodes, LL128 still benefits from fine-grained pipelining within a node due to its smaller transmission granularity. This combination of low latency and high throughput makes LL128 well suited for a broad range of message sizes.

However, LL128 comes with stricter hardware requirements. It depends on atomic 128-byte writes, which must not be split or reordered by the memory system or interconnect. In systems where such operations are not guaranteed, due to PCIe limitations or other architectural constraints, NCCL disables LL128 to avoid data corruption. Protocol selection is thus influenced not only by message size, but also by system-level capabilities.

\subsection{Protocol Selection and Comparison}

NCCL dynamically selects among the Simple, LL, and LL128 protocols at runtime based on user settings (i.e., \texttt{NCCL\_PROTO}), the collective algorithm, and internal performance heuristics. If not explicitly specified, NCCL uses a tuning model that factors in system topology, GPU architecture, message size, and predefined performance metrics to choose the best algorithm-protocol pair. This selection is constrained by resource availability, such as memory for protocol-specific buffers. Typically, LL/LL128 are chosen for small messages to reduce latency, while Simple is used for larger messages to maximize throughput.

\section{Data-Transfer Methods and Transport Layer}

\begin{table}[t]
\centering
\caption{NCCL Communication Characteristics and Transports}
\label{tab:nccl-comm}
\begin{tabular}{@{}cll@{}}
\toprule
 & \makecell{\textbf{Intra-Node}} & \makecell{\textbf{Inter-Node}} \\
\midrule
\textbf{Transport} &
\setlength{\tabcolsep}{0pt}
\begin{tabular}{ll}
    P2P & \texttt{ p2p.cc}\\
    SHM & \texttt{ shm.cc}\\
    NVLS& \texttt{ nvls.cc}
\end{tabular} & 
\setlength{\tabcolsep}{0pt}
\begin{tabular}{ll}
\multirow{2}{*}{NET} & \texttt{ net\_ib.cc}     \\
                     & \texttt{ net\_socket.cc} \\
COLLNET              & \texttt{ coll\_net.cc}
\end{tabular} \\
\midrule
\makecell{\textbf{Physical}\\ \textbf{Interconnect}} & 
\makecell[l]{NVLink\\PCIe}& 
\begin{tabular}{l}
    InfiniBand\\
    RoCE\\
    TCP/IP (Socket)
\end{tabular} \\
\midrule
\textbf{Optimizations} & 
\begin{tabular}{@{}l@{}}
    GPUDirect P2P\\
    \texttt{P2P\_DIRECT}
\end{tabular} & 
\begin{tabular}{@{}l@{}}
    GPUDirect RDMA
\end{tabular} \\
\bottomrule
\end{tabular}
\vspace{-1em}
\end{table}

Efficient data movement is central to NCCL's communication performance, particularly in multi-GPU and multi-node environments. As summarized in Table~\ref{tab:nccl-comm}, NCCL employs distinct data transfer strategies and transport mechanisms depending on whether communication occurs within a single node (intra-node) or across multiple nodes (inter-node), with each transport optimized for specific hardware and interconnect types to support scalable collectives.

\subsection{Intra-node Data Transfer}
\label{sec:intra-node-data-transfer}

\begin{figure}[!htp]
  \centering
  \includegraphics[width=\linewidth]{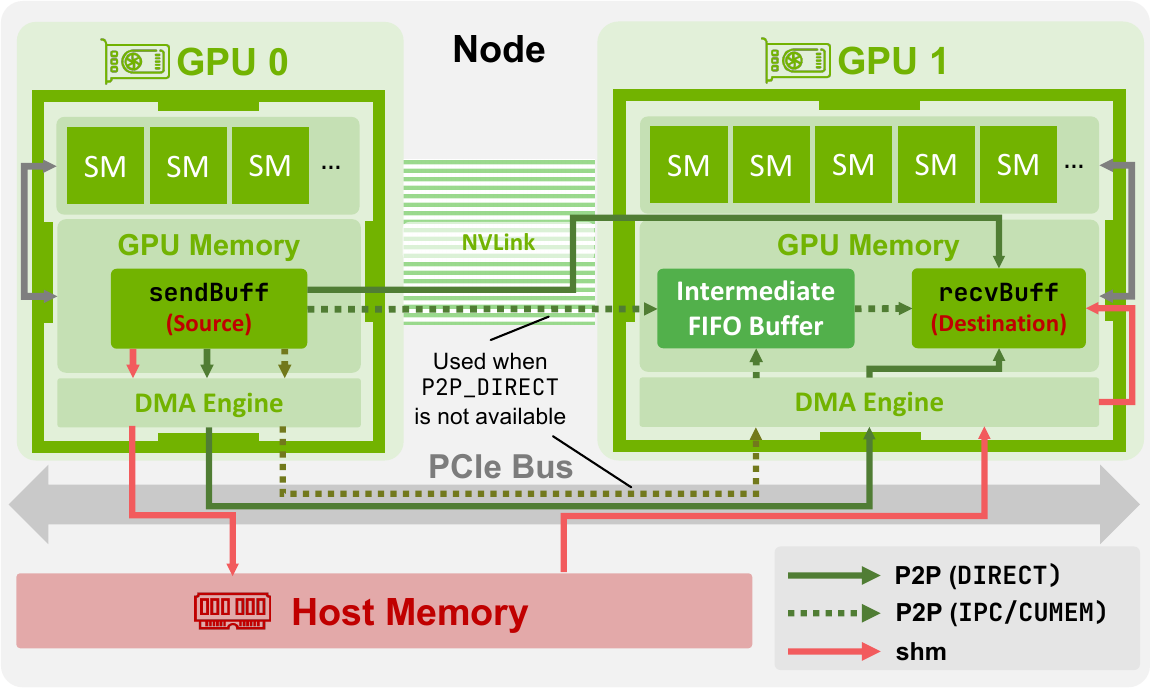}
  \caption{Illustration of intra-node data transfer paths in NCCL. Each path is color-coded to indicate the selected transport and hardware support.}
  \label{fig:intra-node}
  \vspace{-1em}
\end{figure}

NCCL employs a sophisticated and hierarchical approach to intra-node communication, prioritizing the lowest latency and highest bandwidth paths available between GPUs residing on the same physical machine (see Figure~\ref{fig:intra-node}). This strategy heavily leverages NVIDIA's GPUDirect Peer-to-Peer (P2P) technology, which enables GPUs to directly access each other's memory without staging through CPU system memory.

At the core of its intra-node strategy is the P2P transport, primarily managed within \texttt{src/transport/p2p.cc}. When GPUs are interconnected via NVIDIA NVLink, NCCL gives precedence to this path, implementing GPUDirect P2P over NVLink to utilize these dedicated, high-speed, direct GPU-to-GPU links. If NVLink is unavailable,
NCCL can utilize GPUDirect P2P communication over the PCIe bus, also managed by the P2P transport layer. This offers a fallback that is generally much more performant than host-memory staging using \texttt{cudaMemcpy}.

A key optimization in NCCL's P2P transport is the \texttt{P2P\_DIRECT} mode, which is enabled when communicating ranks belong to the same process. While both single-process and multi-process communications utilize GPU-to-GPU transfers without CPU involvement, \texttt{P2P\_DIRECT} mode significantly improves efficiency in two ways. First, it bypasses the need for IPC handles by employing direct GPU memory pointers within the same address space. More importantly, it eliminates an intermediate data copy by using primitives like \texttt{directSend} and \texttt{directRecv}, which transfer data directly between source and destination buffers rather than routing through an \textbf{intermediate FIFO buffer}. Despite this optimized data path, NCCL still maintains correct synchronization using atomic \texttt{head} and \texttt{tail} counters within shared structures (e.g., \texttt{ncclSendMem} and \texttt{ncclRecvMem}) to ensure proper ordering and prevent data races. Thus, \texttt{P2P\_DIRECT} provides substantial performance benefits through both simplified memory addressing and a more direct data transfer path, building upon the foundational GPUDirect P2P capability.

NCCL may leverage the Shared Memory (SHM) transport not only when direct GPU-to-GPU P2P communication is unavailable, but also when P2P is suboptimal. In particular, inter-socket P2P over PCIe often generates P2P packets that are poorly handled by CPUs and result in degraded performance. SHM avoids this by routing traffic through system memory, using PCIe-to-memory and memory-to-PCIe transfers, which CPUs are typically better optimized to process. In SHM mode, one GPU’s controlling process writes data to a shared memory segment, which is then read by the other GPU’s process.

Note that in some multi-socket systems, NCCL may use NICs for intra-node communication between GPUs when each GPU resides on a separate CPU socket with a local NIC supporting GPUDirect RDMA. Rather than traversing the CPU interconnect, NCCL may route data through a GPU–NIC–NIC–GPU path, leveraging PCIe bandwidth to avoid CPU bottlenecks. This behavior is determined by NCCL’s topology-aware logic and can be controlled using environment variables such as \texttt{NCCL\_CROSS\_NIC}.

\subsection{Inter-node Data Transfer}

\begin{figure}[!t]
  \centering
  \includegraphics[width=\linewidth]{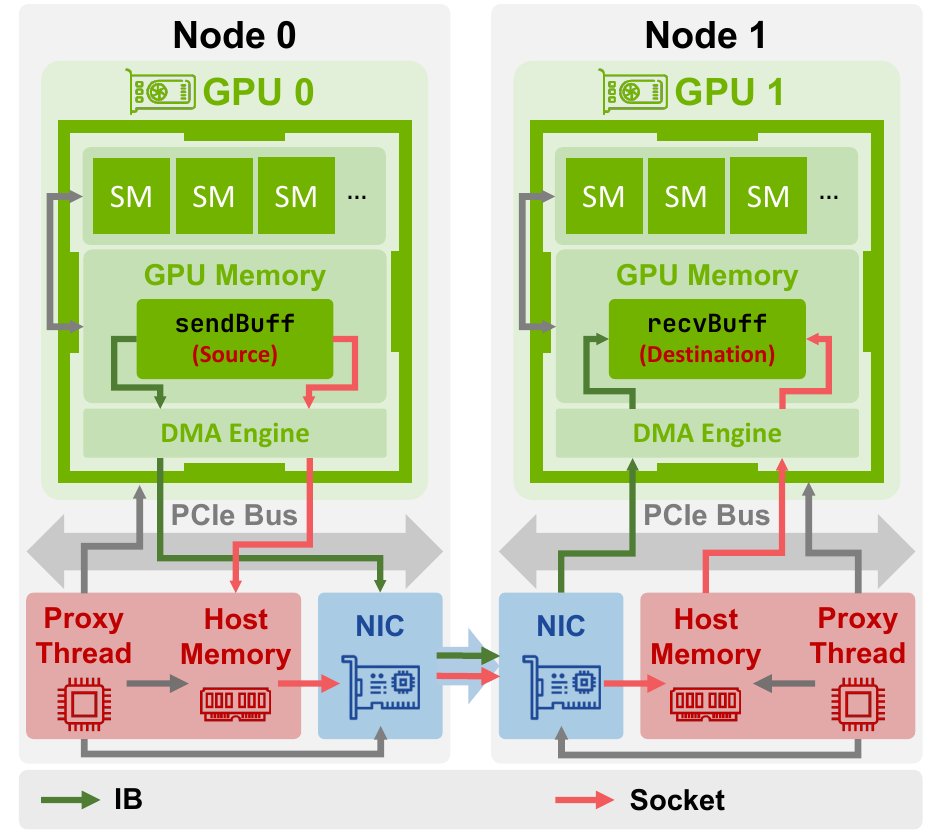}
  \caption{Illustration of inter-node data transfer paths in NCCL. Each path is color-coded to indicate the selected transport and hardware support.}
  \label{fig:inter-node}
  \vspace{-1em}
\end{figure}

Inter-node communication in NCCL orchestrates data exchange between GPUs located in different physical nodes. This process involves the GPU executing NCCL kernels, a proxy thread running on the CPU to manage network operations, and the underlying network fabric. As shown in Figure~\ref{fig:inter-node}, NCCL selects between two primary network transports, namely a standard TCP Socket transport or a high-performance InfiniBand (IB) Verbs transport, based on the available hardware.

\subsubsection{Socket-Based Communication}

When the network interface does not support RDMA, NCCL employs the socket transport, implemented in \texttt{transport/net\_socket.cc}. In this mode, the intermediate buffers are allocated as CUDA pinned memory in the host memory. On the sender side, data is copied from the GPU to this buffer before being sent over the network using standard socket calls. On the receiver side, data is received into the host buffer and then copied to the GPU. This reliance on the host memory as a staging area incurs the overhead of extra memory copies across the PCIe bus. Both sending and receiving follow a rendezvous protocol, where the sender and receiver coordinate buffer readiness before the actual data transfer occurs.

\subsubsection{IB Verbs Transport}

For high-performance networks such as InfiniBand or RoCE, NCCL uses the IB transport, implemented in \texttt{net\_ib.cc}. The IB transport leverages RDMA capabilities to enable direct data movement between nodes with minimal CPU intervention. As with the socket transport, all transfers are staged through an intermediate buffer, but the location of this buffer depends on hardware support and configuration.

By default, if the NIC cannot access GPU memory directly, the intermediate buffer is allocated in host memory. The GPU kernel copies data to this buffer, and the proxy thread posts an RDMA operation to move the data from host memory to the remote node with RDMA write~\cite{nccl_issue_609}. On the receiving side, the process is reversed: the NIC writes incoming data into a host buffer, and the proxy thread coordinates a copy from host to device memory. The proxy thread’s role is to manage these DMA and RDMA operations. As with the socket transport, a rendezvous protocol is used to synchronize the sender and receiver before the data transfer.

In the following paragraphs, we highlight some key features and optimizations implemented in the IB transport.

\paragraph{The GPUDirect RDMA Optimization}

A key optimization in the IB transport is GPUDirect RDMA (GDRDMA), which enables the NIC to access GPU memory directly, eliminating the need for host memory staging. GDRDMA is used only when both the NIC and GPU are connected to the same PCIe switch. In this case, the intermediate buffer is allocated in \emph{GPU memory}. The CPU proxy thread registers this GPU memory with the RDMA-capable NIC using mechanisms such as \texttt{nv\_peer\_mem}\cite{nvidia_gpudirect_rdma_2025} or the Linux DMA-BUF subsystem~\cite{kernel_dma_buf}, allowing the NIC to map and access GPU memory directly. The NIC’s DMA engine then performs RDMA reads or writes directly to or from the GPU, bypassing the CPU and host memory entirely.

\paragraph{Per-peer Multi-channel Connections}
As an optimization to improve bandwidth utilization and reduce congestion, the IB transport instantiates 2 logical channels per remote GPU and per NIC (controlled by the \texttt{NCCL\_IB\_QPS\_PER\_CONNECTION} parameter) by default. Each logical channel maintains its own \texttt{ncclIbSendComm} structure, embedding an independent bundle of InfiniBand QPs. During execution, the host-side network proxy alternates between the two \texttt{sendComm} handles when issuing \texttt{ncclNet->isend()} calls, thereby splitting traffic across the QP sets. This round-robin strategy increases the effective per-QP chunk size, introduces path diversity for ECMP-aware fabrics, and enhances overall interconnect efficiency—all without incurring additional coordination overhead.

\begin{table*}[htbp]
\centering
\begin{threeparttable}
\caption{Supported algorithms and protocols for NCCL collective operations}
\label{tab:collective_support_summary}
\renewcommand{\arraystretch}{1.3}

\begin{tabular}{ccccc}
\toprule
\textbf{AllReduce} & \textbf{Broadcast} & \textbf{Reduce} & \textbf{ReduceScatter} & \textbf{AllGather} \\
\midrule
\begin{tabular}{@{}>{\bfseries}lccc@{}}
Algorithm & Simple & LL & LL128 \\
\midrule
Ring            & \checkmark & \checkmark & \checkmark \\
Tree            & \checkmark & \checkmark & \checkmark \\
CollNet Direct  & \checkmark & \xmark     & \xmark     \\
CollNet Chain   & \checkmark & \xmark     & \xmark     \\
NVLS            & \checkmark & \xmark     & \xmark     \\
NVLS Tree       & \checkmark & \xmark     & \xmark     \\
\end{tabular}
&
\begin{tabular}{@{}ccc@{}}
Simple & LL & LL128 \\
\midrule
\checkmark & \checkmark & \checkmark \\
\xmark     & \xmark     & \xmark     \\
\xmark     & \xmark     & \xmark     \\
\xmark     & \xmark     & \xmark     \\
\xmark     & \xmark     & \xmark     \\
\xmark     & \xmark     & \xmark     \\
\end{tabular}
&
\begin{tabular}{@{}ccc@{}}
Simple & LL & LL128 \\
\midrule
\checkmark & \checkmark & \checkmark \\
\xmark     & \xmark     & \xmark     \\
\xmark     & \xmark     & \xmark     \\
\xmark     & \xmark     & \xmark     \\
\xmark     & \xmark     & \xmark     \\
\xmark     & \xmark     & \xmark     \\
\end{tabular}
&
\begin{tabular}{@{}ccc@{}}
Simple & LL & LL128 \\
\midrule
\checkmark & \checkmark & \checkmark \\
\xmark     & \xmark     & \xmark     \\
\xmark     & \xmark     & \xmark     \\
\xmark     & \xmark     & \xmark     \\
\checkmark & \xmark     & \xmark     \\
\xmark     & \xmark     & \xmark     \\
\end{tabular}
&
\begin{tabular}{@{}ccc@{}}
Simple & LL & LL128 \\
\midrule
\checkmark & \checkmark & \checkmark \\
\xmark     & \xmark     & \xmark     \\
\xmark     & \xmark     & \xmark     \\
\xmark     & \xmark     & \xmark     \\
\checkmark & \xmark     & \xmark     \\
\xmark     & \xmark     & \xmark     \\
\end{tabular}
\\
\bottomrule
\end{tabular}

\vspace{0.2em}
\begin{tablenotes}
\centering
\small
\item \textbf{Legend:} \checkmark = Supported, \xmark = Not supported.
\end{tablenotes}
\end{threeparttable}
\vspace{-1em}
\end{table*}

\paragraph{QP Layout}

For every pair of ranks, the RDMA plugin establishes \emph{two} reliable connection (RC) QPs, one in each direction. The \emph{forward} QP is responsible for the bulk data stream: the proxy issues one or more \texttt{RDMA\_WRITE} work requests that push user data directly into the peer buffer, followed by an \texttt{RDMA\_WRITE\_WITH\_IMM} for completion notification. The behavior of this final request depends on the communication pattern: for small messages, it combines data and size information in a single operation to minimize overhead; for large messages with adaptive routing, it sends a zero-byte write with the transfer size as immediate data to guarantee ordering while allowing the bulk data to use adaptive routing; for aggregated multi-message transfers, it writes precise size information to a remote array while using immediate data for completion signaling. The \emph{reverse} QP carries only a tiny \textbf{clear-to-send (CTS)} message, which is a single \texttt{RDMA\_WRITE} that advertises the remote buffer address, rkeys, and tag information. While the same functionality could theoretically be multiplexed on a single QP, separating the CTS onto its own channel isolates latency-critical control traffic from the bandwidth-hungry data stream, allowing the network to deliver it with minimal head-of-line blocking.

\paragraph{Local Flush with Loop-back \texttt{RDMA\_READ}}
When GPUDirect RDMA is enabled, the sender must ensure that all outstanding PCIe writes reach GPU memory before the kernel consumes the data.  NCCL implements this by issuing a dummy \texttt{RDMA\_READ} after the last receive completes. A dedicated ``flush'' QP is connected to itself, meaning that its ready-to-receive (RTR) stage uses its own local QP number as the destination. Consequently, the read never leaves the host, but the verbs layer still waits for the PCIe completion of prior writes, providing an inexpensive ordering barrier.

\section{NCCL Collective Algorithms}

Collective algorithms are central to NCCL, enabling efficient, synchronized communication between GPUs. They manage data movement and dependencies, optimize communication paths, and scale with increasing GPU counts. NCCL implements these algorithms by breaking each collective operation into low-level communication primitives and distributing them across multiple parallel channels. The choice of algorithm, typically ring or tree, depends on the specific collective operation and relevant execution parameters such as message size and topology \cite{rabenseifner_reductions,patarasuk_yuan,collectives_mpich}. This section outlines the design and main features of NCCL's collective algorithms.

\subsection{Overview of Algorithm and Protocol Support}

While NCCL provides six protocols, not all are applicable to every algorithm, and their availability may vary based on hardware features and runtime constraints.

Table~\ref{tab:collective_support_summary} summarizes the algorithms and communication protocols supported by each of the 5 collective operations in NCCL version 2.19. This information was extracted from the corresponding header files in the \texttt{src/device} directory. In addition to the commonly used Ring and Tree algorithms, the table also highlights support for specialized algorithms, namely CollNet and NVLS. NVLS and CollNet are specialized algorithms primarily designed to optimize AllReduce performance, with NVLS also offering support for ReduceScatter and AllGather by leveraging specific hardware capabilities.

The CollNet algorithms are intended for scenarios where the network infrastructure itself can participate in collective operations, such as using NVIDIA SHARP (Scalable Hierarchical Aggregation and Reduction Protocol) technology, allowing reductions or other partial collective computations to be offloaded to network switches, thereby reducing data movement and latency~\cite{nccl_issue_320}.

CollNet algorithms leverage NVIDIA SHARP (Scalable Hierarchical Aggregation and Reduction Protocol) technology for network-assisted collective operations. CollNet Direct enables all-to-all communication within the node. In contrast, CollNet Chain arranges GPUs linearly, and performs reductions up the chain and broadcasts down~\cite{nccl_issue_919}.

NVLS algorithms are designed to take advantage of NVIDIA's NVLink Switch (NVSwitch) systems, which provide high-bandwidth, direct GPU-to-GPU communication paths within a multi-GPU server or NVSwitch fabric, enabling more efficient collective operations~\cite{nccl_documentation}. Both the plain NVLS and NVLS Tree algorithms use NVLink SHARP for intra-node reduction but differ in inter-node handling: NVLS continues the reduction via CollNet and SHARP-enabled switches, while NVLS Tree uses a tree-based fan-out~\cite{nccl_issue_919}.

However, this paper will not include analyses of NVLS and CollNet, as their implementations rely heavily on specific hardware (NVSwitch and SHARP-enabled networks), making them less representative. We acknowledge that NCCL continues to evolve, recently introducing additional algorithms such as Parallel Aggregated Trees (PAT) in version 2.23~\cite{new_algo_jeaugey, pat_jeaugey}. Nonetheless, as newer algorithms have yet to achieve widespread adoption, our subsequent discussions will remain centered on the Ring and Tree algorithms.

\subsection{Communication Primitives}
\label{sec:communication-primitives}
NCCL implements high-level collective operations by composing them from a set of low-level communication primitives. These primitives form the foundation of NCCL’s collective algorithms, encapsulating basic operations such as sending, receiving, reducing, and copying data across GPUs.

Common primitives include \texttt{send}, \texttt{recv}, \texttt{recvReduceSend}, \texttt{recvCopySend}, and \texttt{recvReduceCopySend}, along with their "direct" variants discussed in Section~\ref{sec:intra-node-data-transfer}. Each primitive represents a distinct data movement or computation pattern, with naming conventions that clearly indicate the sequence of operations. For example, \texttt{recvReduceSend} denotes a step in which a GPU receives data from a peer, performs a reduction with its local buffer, and sends the result to the next GPU. During execution, the NCCL runtime dispatches these primitives iteratively across loop steps, enabling flexible coordination across different algorithms, topologies, and transport layers.

The behavior of each NCCL primitive is further shaped by the selected communication protocol. Synchronization, buffer management, and transfer granularity vary depending on whether the Simple, LL, or LL128 protocol is used. It is important to note that these low-level primitives are heavily optimized for collectives with a fixed, small number of sources and destinations, such as rings and trees, which typically involve one source and one destination (or up to three for certain tree topologies). While this approach enables high efficiency for many standard collective algorithms, it is less effective for patterns like all-to-all, which require handling $N$ sources and $N$ destinations.

\subsection{Iterative Execution of NCCL Collectives}

\begin{table}[htp]
\centering
\caption{NCCL channel buffer sizes for each protocol under the default configuration}
\label{tab:nccl-buffer-pipeline-slots}
\renewcommand{\arraystretch}{1.3}
\begin{tabular}{@{}lccc@{}}
\toprule
\textbf{Protocol} & \makecell{\textbf{Total Channel}\\\textbf{Buffer Size}} & \makecell{\textbf{Buffer Capacity}\\\textbf{per Slot}} & \makecell{\textbf{Effective Data}\\\textbf{per Slot}} \\
\midrule
Simple & 4 MiB & 512 KiB & 512 KiB \\
LL & 256 KiB & 32 KiB & 16 KiB \\
LL128 & $\sim$4800 KiB & 600 KiB & 562.5 KiB \\
\bottomrule
\end{tabular}
\end{table}

\begin{figure}[!t]
  \centering
  \includegraphics[width=\linewidth]{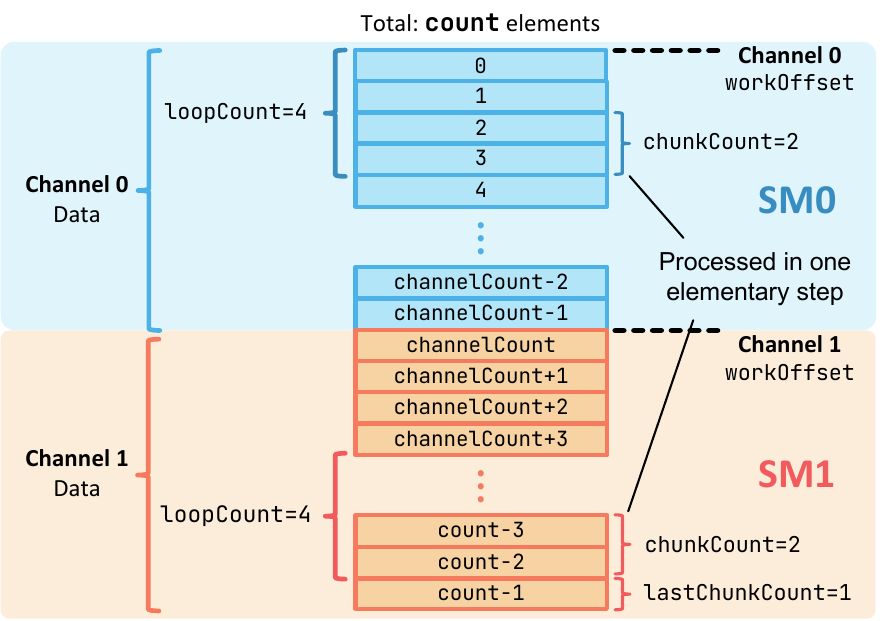}
  \caption{Visualization of NCCL's data partitioning strategy across communication channels and loop iterations.}
  \label{fig:iteration-data-split}
  \vspace{-1em}
\end{figure}

NCCL processes collective operations by first dividing the user’s input data among the available communication channels, enabling parallelism at the channel level. Each channel is responsible for a contiguous segment of the input, determined by the total number of elements (\texttt{count}) and the number of channels. This partitioning is visualized in Figure~\ref{fig:iteration-data-split}, where the total data is split so that each channel, such as Channel 0 and Channel 1, operates independently on its assigned region. The start index for each channel’s work is given by \texttt{workOffset}, and the size by \texttt{channelCount}.

To facilitate efficient data transfer and computation, NCCL allocates a fixed-size buffer for each channel, the capacity of which depends on the chosen communication protocol (Simple, LL, or LL128, as shown in Table~\ref{tab:nccl-buffer-pipeline-slots}). If a channel’s data region is larger than its total buffer, NCCL breaks the data into several \textbf{outer loop iterations}. Each iteration processes a segment of data up to the size of the buffer (\texttt{loopCount} elements per iteration), and the channel cycles through as many loops as needed to cover all its assigned elements.

Within each outer loop iteration, NCCL implements pipelining by dividing the channel buffer into a fixed number of segments, known as \textbf{slots}. It is typically 8, and set by the \texttt{NCCL\_STEPS} parameter. Each slot can independently advance through different stages of communication and computation, allowing the pipeline to overlap data transfers with reduction or copy. During each \textbf{elementary step}, which follows the communication primitives described in Section~\ref{sec:communication-primitives}, a chunk of data (\texttt{chunkCount} elements, or \texttt{lastChunkCount} for the final chunk in a loop) is processed and mapped to the buffer slots. This chunking mechanism allows NCCL to keep the communication channels busy, overlapping new chunks with ongoing operations for maximum throughput.

In NCCL, the basic unit of data movement is called an \textbf{element}, and its meaning depends on the collective operation. For \texttt{ncclAllGather} and \texttt{ncclBroadcast}, each element is a single byte, since these operations focus on efficiently moving and concatenating data. This byte-level granularity gives NCCL flexibility in packing and transferring data, independent of the underlying type. For \texttt{ncclAllReduce}, \texttt{ncclReduceScatter}, and \texttt{ncclReduce}, each element corresponds to the user-defined data type (e.g., \texttt{float} or \texttt{int}), because these operations require arithmetic reductions that are meaningful only at the data type level.

Figure~\ref{fig:iteration-data-split} shows this process in action. Each cell in the figure represents one data element in \texttt{sendBuff}. For illustrative purposes, this example assumes that the number of channels equals 2, \texttt{chunkCount} equals 2, and \texttt{loopCount} equals 4. Channel 0 starts at its \texttt{workOffset} and processes elements in loop iterations of \texttt{loopCount}, breaking them further into chunks of \texttt{chunkCount}. Channel 1 follows the same logic for its own region. By coordinating this partitioning and pipelining, NCCL achieves efficient, parallel, and scalable collective operations across all participating GPUs.

\subsection{Mapping Communication Pipelines to CUDA Hierarchy}

To fully understand NCCL's performance characteristics, it is crucial to examine how these communication channels and data pipelines map onto the GPU's parallel execution model. NCCL orchestrates hundreds of thousands of threads across multiple SMs through a carefully structured hierarchy that aligns with CUDA's thread organization.

\subsubsection{Grid and Block Structure}
NCCL kernels are launched with a grid dimension of \texttt{(nChannels, 1, 1)}, where \texttt{nChannels} represents the number of active communication channels for the operation. This one-to-one mapping ensures that each CUDA block (\texttt{blockIdx.x}) corresponds to exactly one communication channel. Within each block, NCCL uses a variable number of threads, ranging from \texttt{NCCL\_MIN\_NTHREADS} to \texttt{NCCL\_MAX\_NTHREADS}. The exact thread count is determined by NCCL's autotuning system and stored in \texttt{plan->threadPerBlock} during kernel launch preparation.

\subsubsection{Channel-to-Block ID Mapping}
The mapping between \texttt{blockIdx.x} and channel ID is managed through a bitmask-based approach. Each kernel receives a \texttt{channelMask} that specifies which channels are active for the current operation. The device code computes the correspondence between block index and channel ID using bit-population counts: for a given \texttt{blockIdx.x}, the channel ID is determined by finding the \texttt{blockIdx.x}-th set bit in the \texttt{channelMask}.

\subsubsection{Warp-Level Organization}
Within each block, NCCL organizes threads into specialized roles at the warp level. The first two warps handle initialization tasks: warp 0 loads the communicator metadata (\texttt{ncclDevComm}) into shared memory, while warp 1 loads the channel-specific data (\texttt{ncclDevChannel}). The remaining warps perform the actual communication and computation work.

For collective operations, the number of working warps is controlled by the \texttt{nWarps} field in the work descriptor. Different algorithms allocate these warps differently. For example, in NVLS AllReduce, warps are subdivided into groups for different phases. For point-to-point operations, warps are partitioned between send and receive operations, with the distribution computed dynamically based on the number of concurrent transfers.

\subsubsection{Slot-Based Pipeline Execution}
Within each channel's buffer, the \texttt{NCCL\_STEPS} slots enable fine-grained pipelining at the thread level. Threads within a warp collaborate to move data through these slots in a circular manner. Each slot contains a \texttt{ncclConnFifo} structure with fields for mode, offset, size, and data pointer, allowing data to exist in different pipeline states simultaneously: being computed, queued for transmission, in-flight over the network, or ready for consumption.

\subsubsection{Thread-Level Data Movement}
At the finest granularity, NCCL distributes work among individual threads within each warp. For bulk data movement, threads process multiple data elements per iteration in an unrolled fashion. The exact amount depends on the protocol.

Thread divergence is carefully managed through warp-uniform operations. For instance, all threads in a warp perform the same sequence of operations (send, reduce, copy) but on different data elements or memory addresses. This ensures maximum utilization of the GPU's SIMT architecture.

\subsubsection{Concurrent Pipeline Execution}
Critically, NCCL never executes "one communication task" as might be suggested by high-level descriptions. Instead, it maintains dozens of concurrent data movement pipelines simultaneously. Multiple channels execute in parallel across different SMs, multiple slots within each channel pipeline different stages of data flow, and multiple warps within each channel handle different stages of the communication. This multi-level parallelism is essential for achieving high bandwidth utilization.

\begin{figure*}[!t]
    \centering
    \includegraphics[width=1\linewidth]{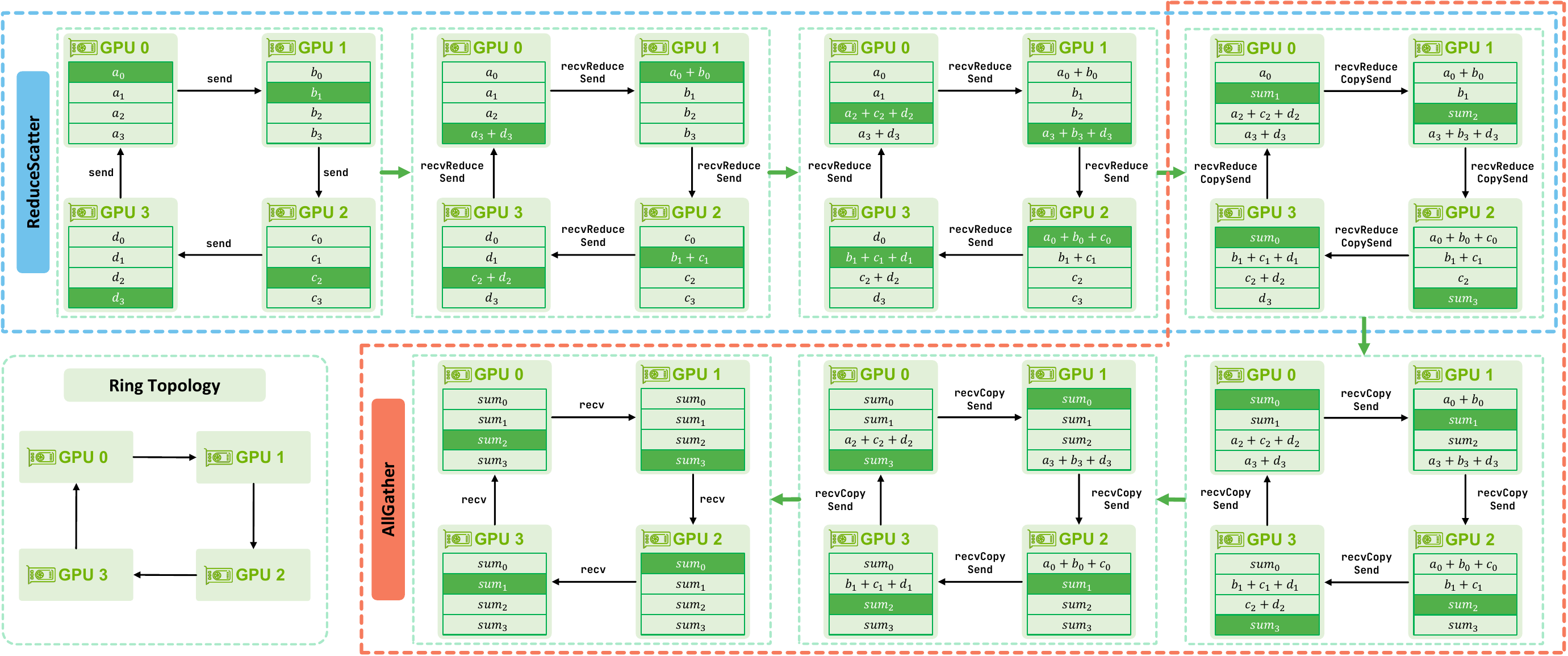}
    \caption{Illustration of the Ring AllReduce algorithm in NCCL across 4 GPUs connected in a ring topology, highlighting the sequence of GPU communication primitives within a single loop iteration.}
    \label{fig:nccl-allreduce-ring}
    \vspace{-1em}
\end{figure*}

\subsection{Qualitative Algorithm Analysis}

Now that we have established that all common NCCL collective algorithms follow an iterative processing model, an important difference lies in whether GPUs can pipeline consecutive loop iterations. Based on this characteristic, algorithms can be grouped into two categories: \textbf{pipelined} and \textbf{non-pipelined}. In the following sections, we organize the collective algorithms accordingly and provide a qualitative analysis of each. For every algorithm, we describe the specific sequence of elementary steps executed within each loop iteration.

While we initially considered a quantitative complexity analysis that would bound algorithm runtimes in terms of parameters such as data size and the alpha-beta model, we found this approach impractical because of the large number of factors that influence performance. Variables such as how GPUs are distributed across nodes have a major effect. For example, 4 GPUs on a single node experience very different bandwidth and latency compared to 4 GPUs placed on separate nodes. Including all these variables would make the model too complex and contradict the goal of keeping complexity bounds simple and useful. For this reason, our analysis remains qualitative and focuses on the essential behaviors rather than attempting to provide detailed theoretical runtime estimates.

\subsubsection{Non-pipelined Pattern}

In the non-pipelined pattern, each GPU must complete all tasks in one iteration before starting the next. Ring AllReduce, Ring AllGather, and Ring ReduceScatter follow this pattern. In the analysis below, $k$ denotes the number of GPUs participating in the collective.

\paragraph{Ring AllReduce}

The Ring AllReduce algorithm in NCCL combines a distributed reduction phase with a data dissemination phase to ensure all $k$ participating GPUs receive the complete, element-wise reduced result. The operation is divided into $2k-1$ steps per loop, as detailed in Table~\ref{tab:ring-allreduce-steps}.

\begin{table}[!htp]
\centering
\caption{Steps in one loop iteration of NCCL Ring AllReduce}
\label{tab:ring-allreduce-steps}
\begin{tabular}{@{}ll@{}}
\toprule
\textbf{Step Index} & \textbf{NCCL Primitive} \\
\midrule
\(0\)                    & \texttt{send} \\
\(1\) to \(k-2\)         & \texttt{recvReduceSend} \\
\(k-1\)                  & \texttt{recvReduceCopySend} \\
\(k\) to \(2k-3\)        & \texttt{recvCopySend} \\
\(2k-2\)                 & \texttt{recv} \\
\bottomrule
\end{tabular}
\end{table}

The Ring AllReduce algorithm begins with a ReduceScatter-like phase, illustrated in the upper portion of Figure~\ref{fig:nccl-allreduce-ring}. Initially, in Step 0, each GPU sends one segment of its local data to its neighbor. In the next $k-2$ steps, each GPU repeatedly executes a \texttt{recvReduceSend} operation: it receives a data segment from its preceding neighbor, performs an element-wise reduction with the corresponding segment of its local data, and forwards the reduced result to the subsequent GPU in the ring. This iterative reduction continues until Step $k-1$. At this step, each GPU receives a data segment, performs a final reduction, thereby producing the fully reduced segment, and copies the result into its designated location within the output buffer before sending this segment onward.

At this step, each GPU receives a data segment, performs a final reduction, and copies the result into its designated location within the output buffer before sending this fully reduced segment onward. For the next $k-2$ steps, each GPU executes a series of \texttt{recvCopySend} operations. In each step, a GPU receives a fully reduced segment from its preceding neighbor, copies it directly into the appropriate position in its output buffer, and forwards this segment unchanged to the next GPU. The Ring AllReduce operation concludes at Step $2k-2$, with each GPU performing a final \texttt{recv} to complete the collection of fully reduced data.

\begin{figure*}[!t]
    \centering
    \includegraphics[width=0.85\linewidth]{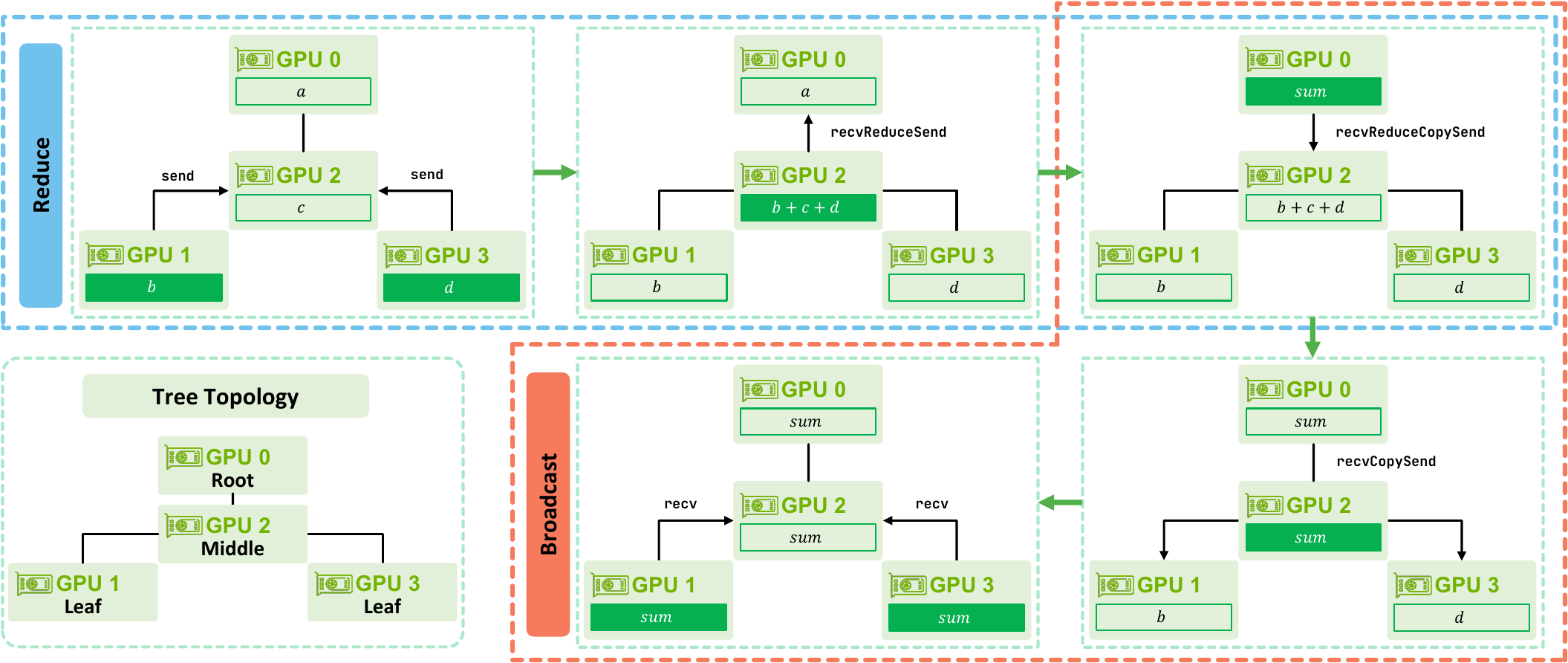}
    \caption{Illustration of the Tree AllReduce algorithm in NCCL across 4 GPUs connected in a tree topology, highlighting the sequence of GPU communication primitives within a single loop iteration.}
    \label{fig:nccl-allreduce-tree}
\end{figure*}

\paragraph{Ring AllGather}
The Ring AllGather algorithm enables each of the $k$ participating GPUs to collect a complete set of data blocks contributed by all ranks. The algorithm proceeds over $k-1$ communication steps using a logical ring topology that connects the GPUs.

In the initial step (Step 0 in Table~\ref{tab:ring-allgather-steps}), each GPU $i$ prepares its local data block. If the operation is in-place, the block is already located in the $i$th segment of the output buffer. Otherwise, the GPU copies the data from its input buffer into that segment using the \texttt{copySend} primitive. After this setup, each GPU sends its local block to its right-hand neighbor.

Over the next $k-2$ steps, each GPU performs a sequence of \texttt{recvCopySend} operations. In each step, a GPU receives a block from its left-hand neighbor, stores it in the correct segment of the output buffer, and forwards it to the right-hand neighbor. The final step is a \texttt{recv} operation that delivers the last missing block. After this step, all GPUs hold a complete, ordered copy of the collective data.

\begin{table}[htp]
    \vspace{-0.5em}
    \centering
    \caption{\textbf{Steps in one loop iteration of NCCL Ring AllGather}}
    \label{tab:ring-allgather-steps}
    \begin{tabular}{ll}
        \toprule
        \textbf{Step Index} & \textbf{Primitives} \\
        \midrule
        0 & \texttt{send} (In-place operation) or \texttt{copySend} \\
        1 to \(k - 2\) & \texttt{recvCopySend} \\
        \(k - 1\) & \texttt{recv} \\
        \bottomrule
    \end{tabular}
\end{table}

\paragraph{Ring ReduceScatter}

The Ring ReduceScatter algorithm performs an element-wise reduction across data blocks initially distributed over $k$ GPUs, followed by scattering unique segments of the fully reduced result back to each GPU. At the beginning, each GPU's \texttt{sendbuff} contains $k$ distinct data blocks, which are progressively reduced as they move around a logical ring topology.

Table~\ref{tab:ring-reducescatter-steps} summarizes the primitives executed in each step of a single loop iteration of Ring ReduceScatter. In the initial step, each GPU $i$ sends one of its local data blocks to its immediate neighbor GPU $(i+1)\%k$, initiating data movement around the ring. During the subsequent $k-2$ steps, each GPU performs a series of \texttt{recvReduceSend} operations: it receives a partially reduced data block from its left neighbor (GPU $(i-1)\%k$), combines this block element-wise with its corresponding local block stored in \texttt{sendbuff}, and sends a different partially reduced block onward to its right neighbor. In the final step, each GPU receives one last data block from its left neighbor, applies the final reduction operation, and copies the fully reduced result directly into its own \texttt{recvbuff}.

\begin{table}[!htp]
    \vspace{-1em}
    \centering
    \caption{\textbf{Steps in one loop iteration of NCCL Ring ReduceScatter}}
    \label{tab:ring-reducescatter-steps}
    \begin{tabular}{ll}
        \toprule
        \textbf{Step Index} & \textbf{Primitives} \\
        \midrule
        Step 0 & \texttt{send} \\
        Step 1 to Step \(k - 2\) & \texttt{recvReduceSend} \\
        Step \(k - 1\) & \texttt{recvReduceCopy} \\
        \bottomrule
    \end{tabular}
\end{table}

\subsubsection{Pipelined Pattern}

The Tree AllReduce, Ring Broadcast, and Ring Reduce algorithms in NCCL follow a pipelined execution pattern.

\paragraph{Tree AllReduce}

The Tree AllReduce algorithm proceeds in two distinct phases within each loop iteration: a \textbf{Reduce} phase followed by a \textbf{Broadcast} phase. The data movement is illustrated by an example involving 4 GPUs in Figure~\ref{fig:nccl-allreduce-tree}. Although the illustration shows a complete tree over four ranks, it is important to note that the branching structure is built only across nodes. Inside each node, NCCL links the local GPUs in a simple chain. In an alternative implementation in NCCL, these two phases are often executed concurrently by partitioning the SMs into two uneven groups. One group handles the reduction toward the root, while the other simultaneously performs the broadcast from the root. This asymmetric allocation dedicates more threads to the bandwidth-intensive reduction phase, enabling better utilization of available resources.

In the Reduce phase, leaf GPUs initiate the reduction by sending their local data upward to their parent using a \texttt{send} operation. Middle GPUs receive data from one or more children using the \texttt{recvReduceSend} primitive, perform element-wise reduction with their own data, and pass the result upward. Finally, the root GPU performs a \texttt{recvReduceCopySend}, completing the reduction by combining the incoming data with its local buffer and copying the fully reduced result into the user-provided output buffer.

In the Broadcast phase, the fully reduced result is propagated back down the tree. The root sends the result to its children using a \texttt{recvCopySend} operation. Middle GPUs receive the data from their parent, copy it into their own output buffer, and forward it to their children using the same \texttt{recvCopySend} primitive. Leaf GPUs receive the data using a simple \texttt{recv} and copy it into their output buffer.

The sequence of device primitives used by each type of GPU role is summarized in Table~\ref{tab:tree-allreduce-steps}.

\begin{table}[!htp]
    \vspace{-1em}
    \centering
    \caption{Steps in one loop iteration of NCCL Tree AllReduce}
    \label{tab:tree-allreduce-steps}
    \begin{tabular}{ll}
        \toprule
        \textbf{GPU Role} & \textbf{Primitives} \\
        \midrule
        Root   & \texttt{recvReduceCopySend} \\
        Middle & \texttt{recvReduceSend} and then \texttt{recvCopySend} \\
        Leaf   & \texttt{send} and then \texttt{recv} \\
        \bottomrule
    \end{tabular}
\end{table}

\begin{figure*}[!t]
  \centering
  \includegraphics[width=0.95\linewidth]{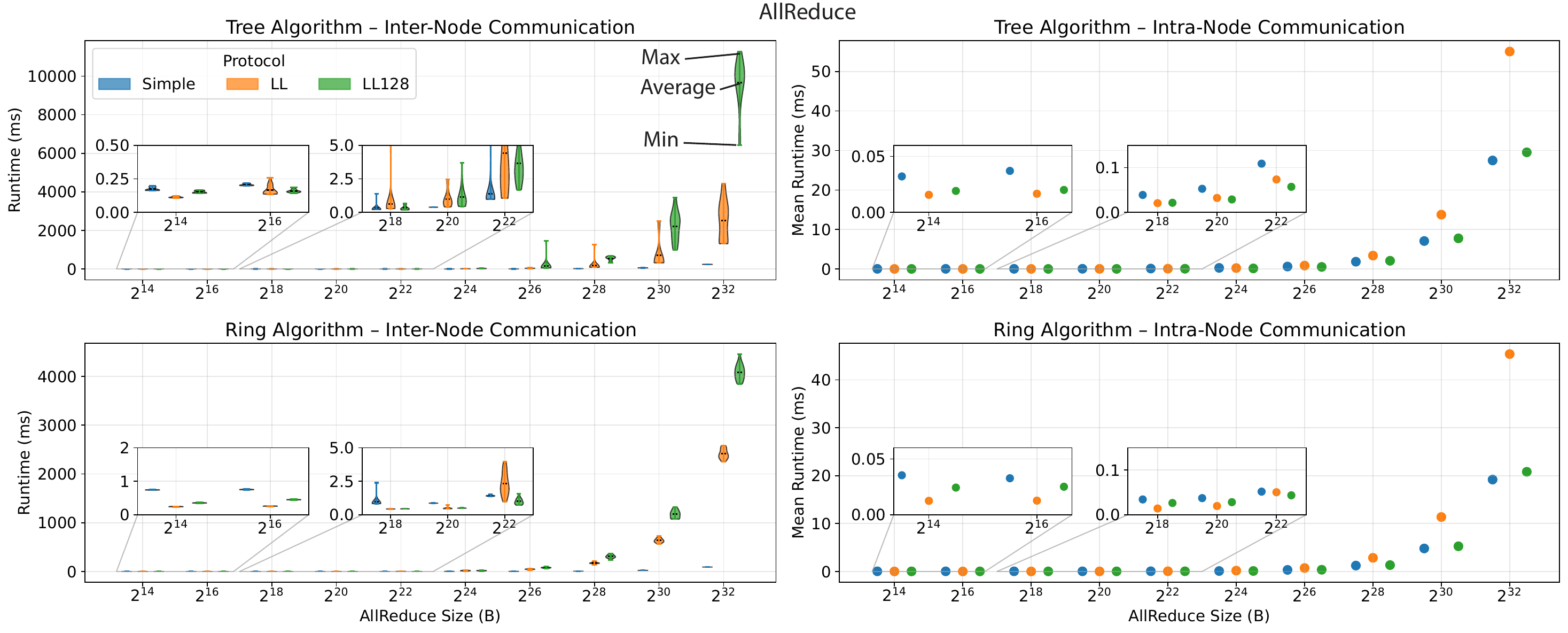}
  \caption{Runtime comparison of protocols for Ring and Tree AllReduce when running inter- and intra-node. Each data point consists of 20 runs with a warm-up phase. For intra-node communication we report only the median value for readability as the variance is very low.}
  \label{fig:allreduce-data}
\end{figure*}

\paragraph{Ring Broadcast}

The NCCL Ring Broadcast algorithm disseminates data from a user-specified root GPU to all other GPUs in the communicator. Although it uses a ring topology, the communication pattern effectively forms a directed chain, beginning at the root and progressing sequentially through each GPU until the last one receives the data.

The operation begins with the root GPU. As indicated in Table~\ref{tab:ring-broadcast-steps}, the root either performs an in-place \texttt{send} operation if its send buffer is also its receive buffer, or a \texttt{copySend} where data from its distinct send buffer is first copied to its receive buffer and then transmitted. In either case, the root sends its data block to its immediate successor in the ring. Each subsequent GPU in the middle of the chain executes a \texttt{recvCopySend} primitive: it receives the data block from its predecessor, copies it into its own receive buffer, and then forwards the data block to its successor. This process continues until the data reaches the last GPU in the chain. This last GPU simply performs a \texttt{recv} operation, copying the incoming data into its receive buffer, and does not send further, as all GPUs in the logical chain have now received the broadcast data.

\begin{table}[!htp]
    \vspace{-1em}
    \centering
    \caption{Steps in one loop iteration of NCCL Ring Broadcast}
    \label{tab:ring-broadcast-steps}
    \begin{tabular}{ll}
        \toprule
        \textbf{GPU Role} & \textbf{Primitives} \\
        \midrule
        Root   & \texttt{send} (in-place) or \texttt{copySend} \\
        Middle & \texttt{recvCopySend} \\
        Last   & \texttt{recv} \\
        \bottomrule
    \end{tabular}
\end{table}

\paragraph{Ring Reduce}

The NCCL Ring Reduce algorithm performs an element-wise reduction of data distributed across multiple GPUs, aggregating the final result onto a user-defined root GPU. Like Ring Broadcast, the operation leverages a logical chain derived from the ring topology, along which data flows and accumulates toward the root.

As shown in Table~\ref{tab:ring-reduce-steps}, the chain begins with the first GPU sending its local data block to the next GPU in the ring. Intermediate GPUs perform the \texttt{recvReduceSend} primitive: each receives a partially reduced block, applies an element-wise reduction using its own corresponding data, and forwards the updated result to the next GPU. This process repeats until the data reaches the destination root. The root GPU completes the operation with a \texttt{recvReduceCopy}: it receives the final partial result, reduces it with its local data, and stores the fully reduced output in its receive buffer.

\begin{table}[!htp]
    \vspace{-1em}
    \centering
    \caption{Steps in one loop iteration of NCCL Ring Reduce}
    \label{tab:ring-reduce-steps}
    \begin{tabular}{ll}
        \toprule
        \textbf{GPU Role} & \textbf{Primitives} \\
        \midrule
        Initiator & \texttt{send} \\
        Middle    & \texttt{recvReduceSend} \\
        Root      & \texttt{recvReduceCopy} \\
        \bottomrule
    \end{tabular}
    \vspace{-1em}
\end{table}

\subsection{Benchmarking}

In this section, we present benchmarking results for NCCL collectives. Figure~\ref{fig:allreduce-data} highlights the runtime performance of the three NCCL communication protocols for AllReduce in both intra-node and inter-node settings, offering a representative view of their behavior. Experiments were conducted on the Alps supercomputing system at the Swiss National Supercomputing Center (CSCS), using 16 nodes equipped with NVIDIA Grace Hopper Superchips (GH200). Each node provides a 150GB/s high-bandwidth intra-node interconnect and connects to the Cray Slingshot interconnect via a 25GB/s per-direction network link~\cite{cscs_alps, understanding_gh200_fusco}.

In the inter-node setting, for both Tree and Ring algorithms, LL and LL128 perform best for small messages (less than 64 KiB). However, as the AllReduce message size increases to the gigabyte range across 16 nodes, their performance drops sharply compared to the Simple protocol. This is mainly due to the overhead of fine-grained, flag-based synchronization in LL and LL128, which requires handling millions of small sync operations (one per 8 or 128 bytes) across the network. While LL128 benefits from larger buffer sizes and is highly efficient over NVLink, these advantages are outweighed by the cumulative synchronization cost over RoCE for large, inter-node transfers. LL128 can even lag behind LL because the extra cost per 128-byte operation becomes significant at scale, or because stalls affect larger data units more under heavy contention. By contrast, the Simple protocol uses much larger transfers with fewer synchronization events, making it less sensitive to network latency and more effective at sustaining high throughput for very large messages.

On the other hand, in the intra-node setting, the LL128 protocol shines with consistent performance across all the message sizes thanks to its ability to fully take advantage of the NVLink connection. In particular, for small messages, LL128 performs as well or only slightly worse than LL while almost matching the performance of Simple (it is 5\% slower than Simple, as expected from Table~\ref{tab:protocol_comparison}) at large messages. The remaining two protocols, LL and Simple perform their best at the opposite extremes, with Simple providing the best performance for large messages and LL for small messages. Finally, we observe that in both intra- and inter-node settings, the Ring algorithm excels for large messages, whereas the Tree algorithm performs best for smaller messages.

There are three takeaways from this benchmarking experiment. First, the results confirm expectations: LL and LL128 are best suited for small messages, especially for inter-node communication, while the Simple protocol consistently outperforms the others for large, distributed transfers. Second, it is important to consider whether the communication is intra-node or inter-node, as different transport algorithms, particularly LL128, exhibit noticeably different performance across these configurations. Finally, while manual protocol selection can be useful for targeted tuning, in most cases it is beneficial to rely on NCCL’s autotuning. Allowing NCCL to select the protocol based on workload characteristics generally provides robust performance and scalability across most use cases.

In addition to AllReduce, we also benchmarked the other collective algorithms. As their behavior follows the same trends observed in AllReduce, we present their runtime results in Figure~\ref{fig:benchmark-data} in the appendix.

\section{Integration into ATLAHS}

Our deep analysis of NCCL’s internal communication patterns, algorithms, and pipelined processing modes has significantly guided the design and capabilities of the ATLAHS toolchain~\cite{atlash_shen}. By characterizing the primitives, data dependencies, and timing behaviors of NCCL operations down to their iterative execution across CUDA streams and communication channels, we were able to accurately decompose collective communication into fine-grained computation, send, and receive events. This knowledge was instrumental in ATLAHS’s GOAL schedule generation process~\cite{goal_hoefler}. Moreover, understanding pipelined vs. non-pipelined collectives allowed us to faithfully model concurrency and overlap, essential for simulating large-scale LLM training.

This NCCL-informed modeling approach allows ATLAHS to accurately emulate GPU communication behavior in real AI training workloads. Unlike prior simulators that largely rely on synthetic patterns or abstract models, ATLAHS captures the execution logic of collective operations with high fidelity. By embedding this insight into the GOAL schedule generator, ATLAHS supports a wide range of topologies and configurations while maintaining simulation errors \textbf{below 5\%}. As shown in our validation and case studies, this design enables ATLAHS to outperform state-of-the-art tools like AstraSim~\cite{astra-sim2_won} in runtime prediction in large-scale multi-GPU environments.

\section{Related Work and Outlook}

Recent studies have offered detailed analyses and performance evaluations of collective communication libraries such as NCCL, MPI, and Gloo in distributed deep learning and HPC environments. For instance, Lee and Lee~\cite{collective_lee} conducted an empirical study comparing these libraries under various training architectures and deployment settings, highlighting NCCL's clear advantage in intra-node GPU-to-GPU communication, particularly for large-scale All-Reduce operations, but also pointing out its performance degradation under virtualization or containerization overheads. Other works, such as the comprehensive survey by Weingram et al.~\cite{xccl_weingram} provide a broad perspective on the ecosystem of collective libraries, reviewing industry solutions like NCCL, RCCL, oneCCL, and Gloo, and noting that while NCCL remains the gold standard for GPU-centric collectives, alternative libraries are rapidly evolving to match its optimizations and hardware support. While studies provide valuable insights into performance and architectural choices, most focus on empirical benchmarks, high-level comparisons, or specific algorithmic innovations. Our work differs by providing an in-depth, systematic analysis of the internal iterative execution algorithms, communication protocols, and data dependencies in collective implementations.

Despite its wide adoption, NCCL faces pressure from recent advances emphasizing adaptability, topology awareness, and fault tolerance. Emerging libraries like Blink achieve notable speedups by dynamically building multiple trees and exploiting advanced network topologies, outperforming NCCL's ring and tree algorithms in large-scale and heterogeneous clusters~\cite{xccl_weingram, blink_wang}. Automated frameworks such as SCCL~\cite{sccl_cai} further push the frontier by synthesizing and tuning collectives for specific hardware, surpassing hand-optimized routines. As distributed AI workloads become more resource-intensive and long-running, fault tolerance and resilience have also become critical requirements~\cite{xccl_weingram}. To keep pace, we believe that future versions of NCCL will need to support automated algorithm selection, robust failure handling, and tighter integration with next-generation fabrics that offer features like in-network computation and smart NICs. Such enhancements are essential for sustaining high performance, scalability, and reliability in ever-more demanding distributed training environments.

\section{Conclusion}

This paper presents a systematic and in-depth analysis of the NVIDIA Collective Communication Library (NCCL). Our investigation examined NCCL’s communication protocols, emphasizing their design trade-offs and dynamic selection logic, as well as the data transfer mechanisms employed in both intra-node and inter-node settings. We also analyzed NCCL’s widely used ring and tree-based collective algorithms, detailing their iterative execution models and the sequence of GPU communication primitives they employ. These insights go beyond academic interest as they are foundational to the development of ATLAHS, an application-trace-driven network simulation toolchain capable of accurately modeling the communication behavior of large-scale AI training workloads. By uncovering the internal mechanisms and performance-critical decisions in NCCL, this work provides system researchers, network architects, and performance engineers with the insights needed to diagnose bottlenecks, optimize communication patterns, and inform the design of future high-performance collective libraries.

\section{Acknowledgment}
The authors would like to thank Tiancheng Chen for his helpful suggestions. This project has received funding from the European Research Council (ERC) under the European Union’s Horizon 2020 program (grant agreement PSAP, No. 101002047). We also thank the Swiss National Supercomputing Center (CSCS) for supporting this project and providing the computational resources used in this work. The authors used ChatGPT-4o and 4.5 to assist with light editing and proofreading throughout the manuscript. All content and ideas are the original work of the authors.

\bibliographystyle{ieeetr}
\bibliography{references}

\appendix

\begin{figure*}
  \centering
  \includegraphics[width=\linewidth]{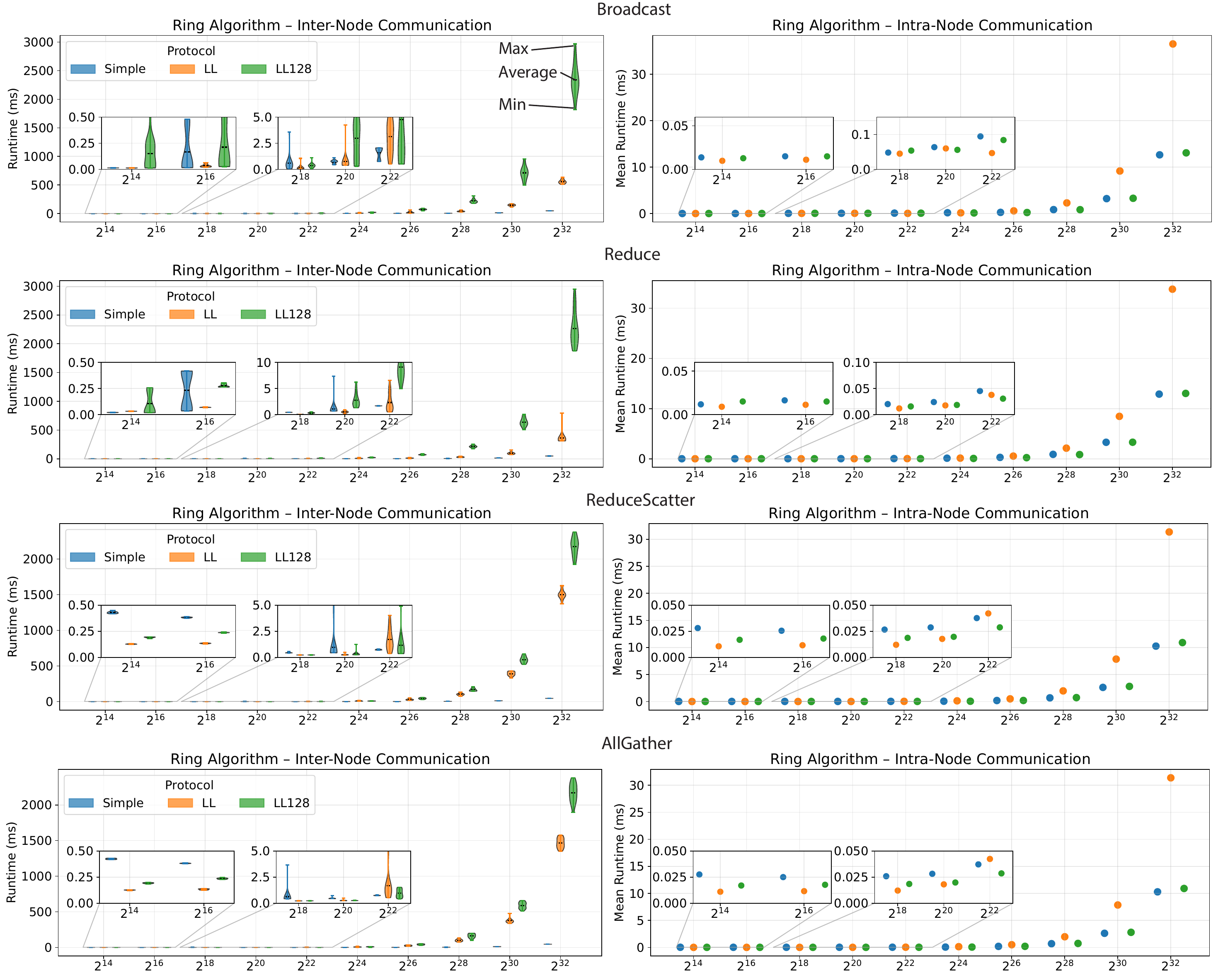}
  \caption{Runtime comparison of protocols for various NCCL collectives when running inter- and intra-node. Each data point consists of 20 runs with a warm-up phase. For intra-node communication we report only the median value for readability, as the variance is very low.}
  \label{fig:benchmark-data}
\end{figure*}

\end{document}